\documentclass[showpacs,amsmath,amssymb,prd]{revtex4}
          
\usepackage{graphicx}
\usepackage{dcolumn}
\usepackage{bm}
\usepackage{epsfig}
\usepackage{amsmath}
\usepackage{color}

\newcommand{\be}{\begin{equation}}
\newcommand{\ee}{\end{equation}}
\newcommand{\ba}{\begin{eqnarray}}
\newcommand{\ea}{\end{eqnarray}}
\def\eps{\epsilon}
\def\veps{\varepsilon}

\def\g{\gamma}
\def\G{\Gamma}
\def\p{\prime}

\begin{document}
\input{epsf}

\title{Long-lived PeV--EeV Neutrinos from GRB Blastwave}

\author{Soebur Razzaque}

\email{srazzaque@uj.ac.za}

\affiliation{Department of Physics, University of Johannesburg, PO Box
  524, Auckland Park 2006, South Africa}

\begin{abstract} 
  Long duration gamma-ray bursts are powerful sources that can
  accelerate particles to ultra-high energies.  Acceleration of
  protons in the forward shock of the highly relativistic GRB
  blastwave allows PeV--EeV neutrino production by photopion
  interactions of ultra-high energy protons with X-ray to optical
  photons of the GRB afterglow emission.  Four different blastwave
  evolution scenarios are considered: adiabatic and fully radiative
  blastwaves in a constant density circumburst medium and in a wind
  environment with the particle density in the wind decreasing
  inversley proportional to the square of the radius from the center
  of the burst.  The duration of the neutrino flux depends on the
  evolution of the blastwave, and can last up to a day in the case of
  an adibatic blastwave in a constant density medium.  Neutrino fluxes
  from the three other blastwave evolution scenarios are also
  calculated.  Diffuse neutrino fluxes calculated using the observed
  rate of long-duration GRBs are consistent with the recent IceCube
  upper limit on the prompt GRB neutrino flux below PeV. The diffuse
  neutrino flux needed to explain the two neutrino events at PeV
  energies recently detected by IceCube can partially come from the
  presented GRB blastwave diffuse fluxes.  Future observations by
  IceCube and upcoming huge radio Askaryan experiments will be able to
  probe the flux models presented here or constrain the GRB blastwave
  properties.
\end{abstract}


\date{\today}
\maketitle

\section{Introduction}

The current paradigm of a long-duration (typical duration $\sim 10$~s)
Gamma-Ray Burst (GRB) is based on core-collapse of a massive ($\gtrsim
30M_\odot$) progenitor star \cite{MacFadyen:1998vz, Woosley:2005gy} to
a blackhole or highly magnetized neutron star (central engine) with
subsequent emission of a short-lived and highly relativistic jetted
ejecta \cite{Meszaros:1993cc}, with a bulk Lorentz factor $\G \gtrsim
100$--$1000$.  Highly variable, on time scales as short as $\Delta
t\sim 10^{-3}$~s, prompt $\g$-ray emission in the keV--MeV range is
thought to originate from the ejecta matreial in the jet itself
\cite{Rees:1994nw}, although the exact emission mechanism is unknown.
Synchrotron radiation by electrons that are accelerated in the shocks
between outflowing materials inside the GRB jet (internal shocks)
and/or thermal radiation from the jet photosphere are plausible
mechanisms to produce the observed keV--MeV $\g$ rays (see, e.g.,
Refs.~\cite{Piran:2004ba, Zhang:2003uk} for reviews).  Protons
co-accelerated with electrons in the internal shocks to Ultra High
Energies (UHE, $\gtrsim 10^{18}$~eV) have been proposed
\cite{Waxman:1995vg} as UHE Cosmic Rays (UHECRs), once escaped from
the GRB jet and observed on the Earth.

Acceleration of UHECRs in the internal shocks leads to TeV--PeV $\nu$
production from interactions of CR protons with the ambient keV--MeV
photons via $p\g$ interactions \cite{Waxman:1997ti}.  The radii from
the central engine at which the internal shocks take place, however,
depend crucially on $\Gamma$.  Optimistic calculations with $\G \sim
100$ and $\Delta t\sim 10^{-3}$~s result in a radius small enough for
the $p\g$ interaction opacity to reach $\sim 1$, thus efficient
production of $\nu$'s in the TeV--PeV range.  No such $\nu$'s,
correlated with GRBs, have been detected by the IceCube Neutrino
Observatory \cite{Abbasi:2012zw} or by the ANTARES neutrino telescope
\cite{Adrian-Martinez:2013sga} from stacking analysis of GRBs which
took place during their respective operations in the last few years.
These results severely constrain the most optimisitc internal-shock
$\nu$ flux models (see, however, Refs.~\cite{Dermer:2003zv,
Murase:2005hy, Hummer:2011ms}).  One possibility, barring a scenario
where GRBs are inefficient accelerators of UHECRs, is that GRBs have
larger $\G$ than used for $p\g$ opacity calculation.  Recent modeling
of GeV $\g$-ray data from the {\it Fermi Gamma Ray Space Telescope}
also reveals that $\G \gg 100$ at least for a large fraction of the
GRBs (see, e.g., Ref.~\cite{Gehrels:2013xd}) detected by its
high-energy ($\gtrsim 20$~MeV--300~GeV) instrument, the Large Area
Telescope (LAT) \cite{Atwood:2009ez}.

The GRB jet drives a blastwave ahead of the ejecta and slows down by
accumulating particles from the ambient medium.  After a time
$t=t_{dec}$ when the kinetic energies of the blastwave and the ejecta
are roughly equal, the blastwave decelerates in a self-similar fashion
\cite{Blandford:1976uq}.  Synchrotron radiation by electrons in the
external forward shock of such a decelerating blastwave has
successfully described multiwavelength observation --- from X rays to
radio --- of GRB afterglows \cite{Meszaros:1996sv, Sari:1997qe}.
Detection of sustained GeV emission by {\it Fermi}-LAT, long after the
GRB prompt emission phase is over and with smoothly decaying flux
characteristcs as observed in X-ray to radio afterglows, from a number
of GRBs provide strong evidence \cite{DePasquale:2009bg,
  Fermi+110731A} that long-lived GeV emission is also part of the
afterglow \cite{Kumar:2009vx, Ghisellini+10}.  This requires
acceleration of electrons to the maximum allowable limit from
synchrotron cooling and often exceeding it \cite{Piran:2010ew,
  Atwood:2013dra}.  A combined electron-proton synchrotron radiation
scenario, during the coasting phase of the GRB fireball
\cite{Razzaque:2009rt} and during the deceleration phase of the blast
wave \cite{Razzaque:2010ku}, may alleviate this problem.

Indeed protons co-accelerated with the electrons in the GRB blastwave
have been suggested to produce UHECRs \cite{Vietri:1995hs}.  These
protons, if interacting with electron-synchrotron photons in the
blastwave, should also produce $\nu$'s via $p\g$ interactions.  In
this work, using analytic and numerical methods, we calculate fluxes
of these $\nu$'s based on different blastwave evolution scenarios.  We
assume $\Gamma \gg 100$ and rapid slow down of the GRB blastwave on a
time scale $\sim 10$~s, as would be required to explain GeV $\g$-ray
emission from the external forward shock in the blastwave.  Note that
our $\nu$ flux model is different from those in
Refs.~\cite{Waxman:1999ai, Dai+01} who calculated fluxes from the
short-lived external reverse shock that propagates into the ejecta
material which may produce optical emission at an earlier stage of the
GRB evolution (see, however, Ref.~\cite{Murase:2007yt} for a
long-lived $\nu$ flux emission model from the reverse shock in
connection with shallow-decay X-ray light curve after the prompt
$\gamma$-ray emission \cite{Nousek:2005fm, Zhang:2005fa}).  The $\nu$
fluxes that we calculate last for a longer time, albeit with
progressively lower intensity in time.  Earlier work on forward-shock
neutrino emission focused on adiabatic blastwave model in constant
density medium \cite{Dermer:2000yd, Li:2002dw}.  Here we carry out
comprehensive study of four different blastwave evolution models.

The organization of this paper is as follows.  In Sec.\ II we set up
our physical model of the GRB blastwave and target photon field for
$p\g$ interactions.  We calculate $\nu$ fluxes in Sec.\ III from a CR
acceleration scenario and briefly discuss $\nu$ detection prospects in
Sec\ IV.  We discuss our results and draw conclusions in Sec.\ V.  A
number of essential formulas are provided in Appendix \ref{appendA} in
order to calculate the synchrotron photon spectra in the GRB blastwave
which are targets for $p\g$ interactions.  We also give analytic
expressions to calculate $p\g$ opacities and CR parameters in Appendix
~\ref{appendB}.  Some scaling formulas for pion and muon decays are
given in Appendix \ref{appendC}.

\section{$p\g$ interaction in GRB blastwave}

We consider photopion production mechanism and associated chain decay
of charged pion and muon ($\pi^+ \to \mu^+ + \nu_\mu \to e^+ + \nu_e +
{\bar \nu}_\mu + \nu_\mu$ and charge conjugate reactions for $\pi^-$)
for UHE $\nu$ flux calculation from a GRB blastwave.  We assume that
UHECRs are accelerated in the forward shock that propagates into the
blastwave and interact with synchrotron photons from electrons which
are accelerated in the same shock.  The observed synchrotron spectrum
which would constitute the target photons for $p\g$ interactions,
however, depends on the properties of the GRB blastwave and the
surrounding environment.  We discuss this briefly here and refer
interested readers to Refs.~\cite{Sari:1997qe, Chevalier+00,
  Granot:2001ge, Panaitescu:2001fv, Ghisellini+10} for further
details.

\subsection{Blastwave models and synchrotron flux}

Given an isotropic-equivalent kinetic energy $E_{k}=10^{55}
E_{55}$~erg and an inital bulk Lorentz factor $\Gamma_0 = 10^{2.5}
\Gamma_{2.5}$, the GRB ejecta (fireball) in the Inter-Steller Medium
(ISM) of uniform density $n(R) = n_0$~cm$^{-3}$; where $R$ is the
distance from the center of the explosion, decelerates on a time scale
\ba
t_{dec,i} &=& \left[ 
\frac{3E_{k} (1+z)^3}{64\pi nm_pc^5 \Gamma_0^8}
\right]^{1/3}
\nonumber \\ &=&
33.3\, (1+z) n_0^{-1/3} \Gamma_{2.5}^{-8/3} E_{55}^{1/3} ~{\rm s}.
\label{dec_time_ism}
\ea
In case of a wind-type medium with a density profile $n(R) = AR^{-2}$,
the deceleration time scale is
\ba
t_{dec,w} &=& 
\frac{E_{k} (1+z)}{16\pi Am_pc^3 \Gamma_0^4}
\nonumber \\ &=&
1.5\, (1+z) A_\star^{-1} \Gamma_{2.5}^{-4} E_{55} ~{\rm s},
\label{dec_time_wind}
\ea
where $A = {\dot M}_w/(4\pi v_w m_p) = 3.02\times 10^{35}
A_\star$~cm$^{-1}$ with $A_\star \equiv {\dot M}_{-5}/v_{8}$
corresponding to a mass-loss rate of ${\dot M}_w = 10^{-5}{\dot
  M}_{-5}M_\odot$~yr$^{-1}$ in wind, by the progenitor star, with
velocity $v_w = 10^8 v_8$~cm~s$^{-1}$.  After deceleration, the
blastwave driven by the GRB ejecta evolves in self-similar fashion
depending on whether it is adiabatic or radiative. The bulk Lorentz
factor $\G (t)$ of an adiabatic and a fully radiative blastwave
evolves in a constant density ISM as
\ba
\Gamma_{ad,i} (t) &=& \Gamma_0 (t_{dec,i}/4t)^{3/8}; \nonumber \\
\Gamma_{ra,i} (t) &=& \Gamma_0 (t_{dec,i}/7t)^{3/7},
\label{ism_bulk_Lorentz}
\ea
respectively.  In case of a wind-type medium the bulk Lorentz factor
evolves as
\ba
\Gamma_{ad,w} (t) &=& \Gamma_0 (t_{dec,w}/4t)^{1/4}; \nonumber \\
\Gamma_{ra,w} (t) &=& \Gamma_0 (t_{dec,w}/7t)^{1/3},
\label{wind_bulk_Lorentz}
\ea
respectively, for an adiabatic and a fully radiative blastwave.  The
radius of the blastwave correspondingly increases as
\ba
R(t) &=& \frac{2\Gamma^2 (t) act} {1+z};
\label{blastwave_radius}
\ea
after $t=t_{dec}$.  Here $a=4$ and $a=7$ for adiabatic and radiative
blastwave, respectively.  The numerical values of the radius along
with the bulk Lorentz factor in the four different scenarios in
Eqs.~(\ref{ism_bulk_Lorentz}) and (\ref{wind_bulk_Lorentz}) are listed
in Appendix \ref{appendA}.  Note that among the four scenarios, the
blastwave radii satisfy the relation $R_{ad,i} (t) > R_{ra,i} (t) >
R_{ad,w} (t) > R_{ra,w} (t)$ until about 3000~s with all parameters in
Eqs.~(\ref{bw_R_ad_i}), (\ref{bw_R_ra_i}), (\ref{bw_R_ad_w}) and
(\ref{bw_R_ra_w}) equal to unity.

A fraction $\epsilon_B$ of the forward-shock energy is believed to be
converted into magnetic energy with a field strength 
\be 
B^\p (t) = [32\pi \epsilon_B n(R) m_p c^2]^{1/2}\Gamma(t) 
\label{B_field}
\ee 
in the comoving blastwave frame (variables are denoted with primes in
this frame).  The magnetic field for the four different blastwave
evolutions are given in Appendix \ref{appendA}.  Electrons accelerated
in the shock is expected to have three characteristic Lorentz factors:
(i) minimum; (ii) cooling; and (iii) saturation.  These are given
below, respectively, as
\ba
\g^\p_m (t) &=& \epsilon_e \frac{m_p}{m_e} \Gamma(t); \nonumber \\
\g^\p_c (t) &=& \frac{6\pi m_ec(1+z)}
{\sigma_T t B^{\p2}(t)\Gamma(t)}; \nonumber \\
\g^\p_s (t) &=& \left[ \frac{6\pi e}{\sigma_T B^\p(t) \phi} 
\right]^{1/2},
\label{electron_Lorentz}
\ea
where $\epsilon_e$ is the fraction of bulk kinetic energy (mostly in
protons) that is converted to random electron energy, $\sigma_T$ is
the Thomson cross section, and $\phi$ is the number of gyro-radius
needed for electron acceleration in the $B^\p$ field.  These three
electron Lorentz factors correspond to three breaks in the observed
sychrotron spectrum at characteristic synchrotron frequencies
\be
h\nu (t) = \frac{3}{2} \frac{B^\p(t)}{B_Q} \g^{\p2}(t)
m_ec^2 \frac{\Gamma(t)}{1+z},
\label{sync_freq} 
\ee
where $B_Q = 4.41\times 10^{13}$~G.  These frequencies are listed in
Appendix \ref{appendA} for the four scenarios described above.
Another break due to synchrotron self-absorption frequency $\nu_a$
may also appear in the spectrum, which is much below $\nu_c$ and in
radio frequencies.  For our calculations we will restrict ourselves to
the fast-cooling regime valid for $t<t_0$, where $t_0$ is derived from
the condition $\nu_m (t_0) = \nu_c (t_0)$ and are given in Appendix
\ref{appendA}.

The maximum synchrotron flux at $\nu_m$ is given by
\be
F_{\nu,m} = \frac{N_e}{4\pi d_{L}^2} 
\frac{P(\g^\p_m)}{\nu_m} \frac{\Gamma^2}{(1+z)^2}, 
\label{max_flux}
\ee
where $N_e = (4/3)\pi R^3 n$ is the total number of electrons in the
blastwave, $d_L$ is the luminosity distance and
\be
P(\g^\p_m) = \frac{c\sigma_T}{6\pi} B^{\p 2} \g_{m}^{\p 2},
\label{sync_power}
\ee 
is the synchrotron power for electrons with Lorentz factor $\g^\p_m$.
The maximum flux for the four different scenarios are given in
Appendix \ref{appendA} as well.

\subsection{$p\g$ interaction  efficiency}

The proper density of the observed synchrotron photons with flux
$F_\nu$, assuming isotropically distributed in the GRB blaswave frame,
is given by
$$
n^\p_\g (\veps^\p) = \frac{2 d_L^2 (1+z) F_\nu}{R^2 c \Gamma \veps^\p},
$$
where $\veps^\p \equiv h\nu^\p = h\nu (1+z)/\Gamma$.  The flux $F_\nu$
is a broken power law and the break frequencies are given in
Appendix~\ref{appendA} as discussed previously.  The photon spectrum
in the comoving frame, which also evolves with time as does $F_\nu$,
follows from the synchrotron spectrum as
\ba
n'_\gamma (\veps') &=& \frac{2 d_L^2 (1+z) F_{\nu,m}}
{R^2 c \Gamma \veps^\p_m}
\nonumber \\ &\times & 
\begin{cases}
(\veps^\p_c/\veps^\p_m)^{-3/2} (\veps^\p/\veps^\p_c)^{-2/3} 
~;~ \veps^\p_a < \veps^\p < \veps^\p_c \cr
(\veps^\p/\veps^\p_m)^{-3/2} 
~;~ \veps^\p_c \le \veps^\p \le \veps^\p_m \cr
(\veps^\p/\veps^\p_m)^{-k/2-1}
~;~ \veps^\p_m < \veps^\p < \veps^\p_s ~,
\end{cases}
\label{fast_cooling_spectrum}
\ea
in the fast-cooling regime ($t<t_0$), where $k$ is the spectral index,
$n(\g^\p) \propto \g^{\p-k}$, of the shock-accelerated electrons.  We
do not consider the slow-cooling regime ($t>t_0$) since $F_\nu$ and,
as we will see shortly, the cosmic-ray flux decay significantly by the
time $t\sim t_0$.

The scattering rate for $p\gamma$ interactions, in the comoving frame,
with target photons $n^\p_\g (\veps^\p)$ for a proton with Lorentz
factor $\g_p^\p$ is given by
\be
K_{p\g} (\g_p^\p)  = \frac{c}{2\gamma_p^{\p 2}} 
\int_{\veps^\p_{\rm th}}^{\infty} d\veps^\p_r 
\veps^\p_r \sigma_{p\gamma} (\veps^\p_r)
\int_{\frac{\veps^\p_r}{2\gamma^\p_p}}^{\infty} 
d\veps^\p \frac{n^\p_\gamma(\veps^\p)}{\veps^{\p 2}}.
\label{pgamma_rate}
\ee
Here $\veps^\p_r = \gamma^\p_p \veps^\p (1-\beta_p \cos\theta)$ is the
photon energy in the rest frame of the proton with an angle $\theta$
between their directions, $\sigma_{p\gamma}$ is the $p\gamma$
interaction cross section, and $\veps^\p_{\rm th} = m_\pi c^2+
m_\pi^2c^2/2m_p$ is the pion production threshold energy for photons.
We have used two models for the $p\g$ cross section.  The first is
$p\gamma \to n\pi^+$ production via $\Delta(1232)$ resonance, with a
cross-section $\sigma_{\Delta} (\veps^\p_r) = \sigma_0\Gamma_\Delta^2
(s\veps_r^\p)^{-2} [\Gamma_\Delta^2 s + (s - m_\Delta^2)^2 ]^{-1}$,
where $s=m_p^2c^4+2 \veps'_r m_pc^2$, $\sigma_0 = 3.11\times
10^{-29}$~cm$^2$, $\Gamma_\Delta = 0.11$~GeV is the width of the
resonance.  The peak of the cross section is $\sigma_{\rm pk}=
4.12\times 10^{-28}$~cm$^2$ at $\veps'_{r,\rm pk} = 0.3$~GeV.  In the
second case, we have used the full $p\g$ cross section from the SOFIA
code \cite{Mucke:1999yb} that includes additional resonance channels
as well as the direct pion production channels.  We define an optical
depth for $p\g$ interactions as
\be
\tau_{p\g}(\g^\p_p) = K_{p\g} (\g_p^\p) t^\p_{dyn} 
= K_{p\g} (\g^\p_p) \frac{R}{2ac\Gamma},
\label{pg_opacity}
\ee
where $t^\p_{dyn} = t\Gamma/(1+z)$ is the dynamic time from
Eq.~(\ref{blastwave_radius}).

\begin{widetext}

\begin{figure}[t]
\includegraphics[width=3.in]{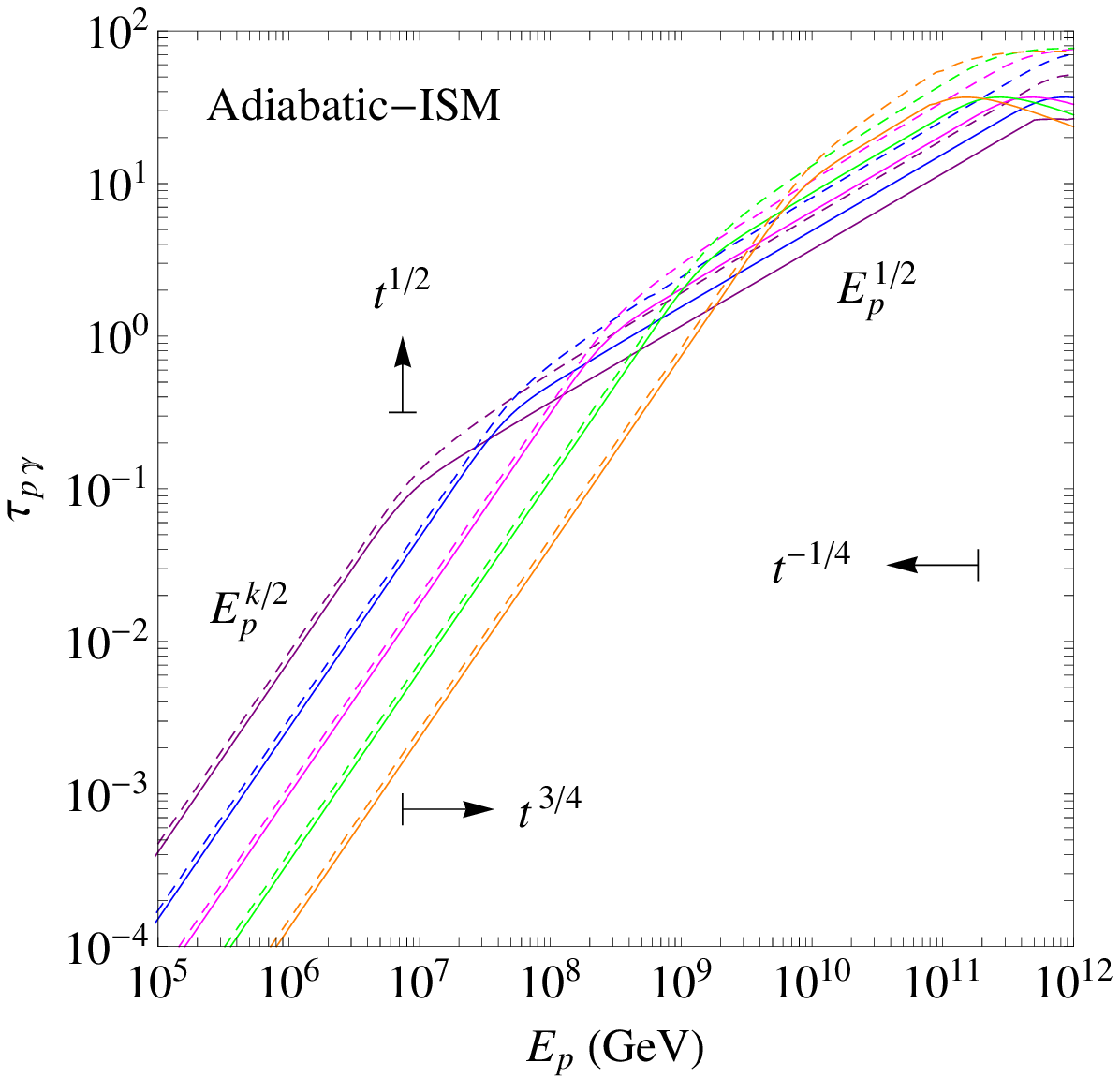}
\includegraphics[width=3.in]{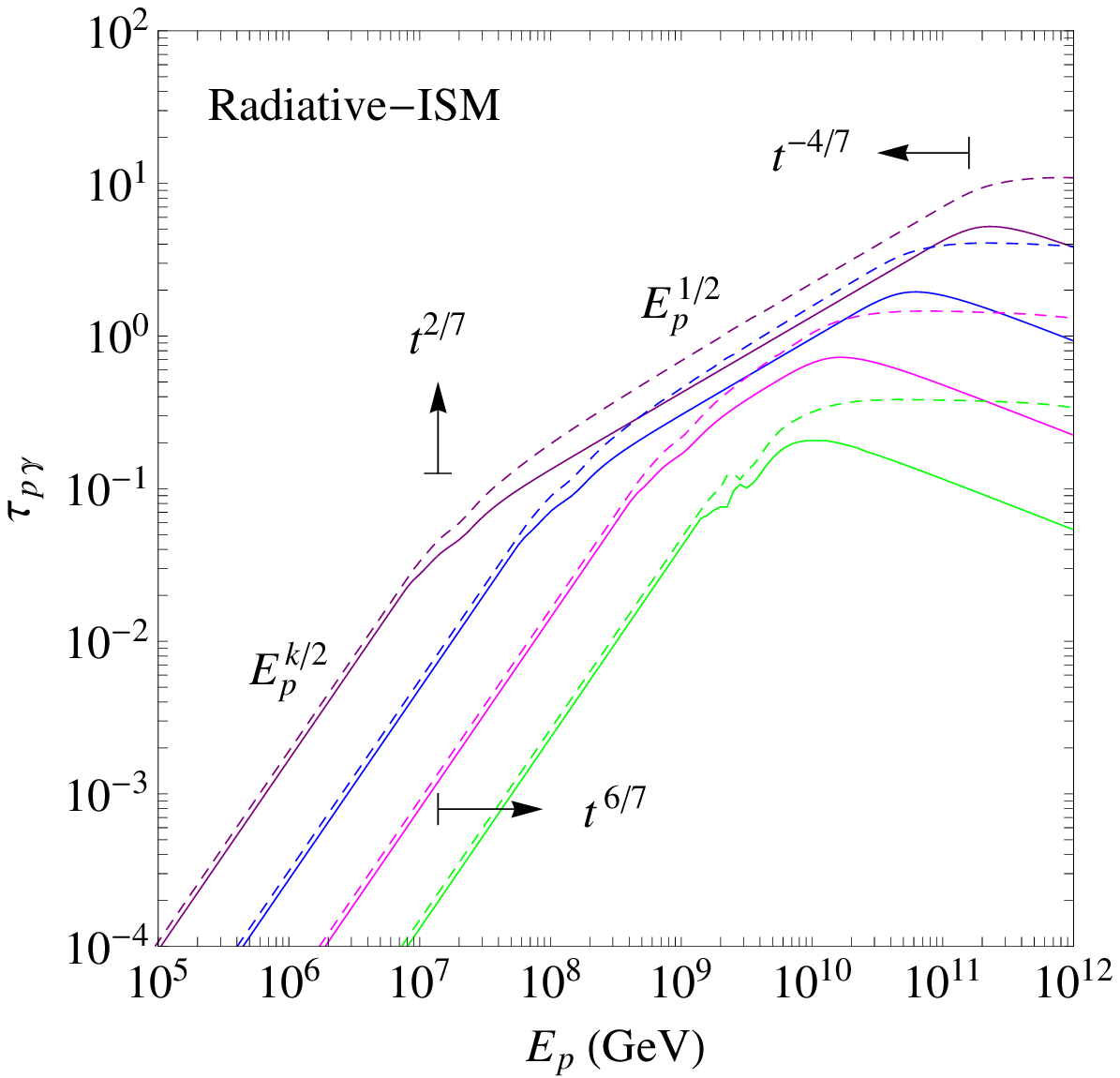}
\includegraphics[width=3.in]{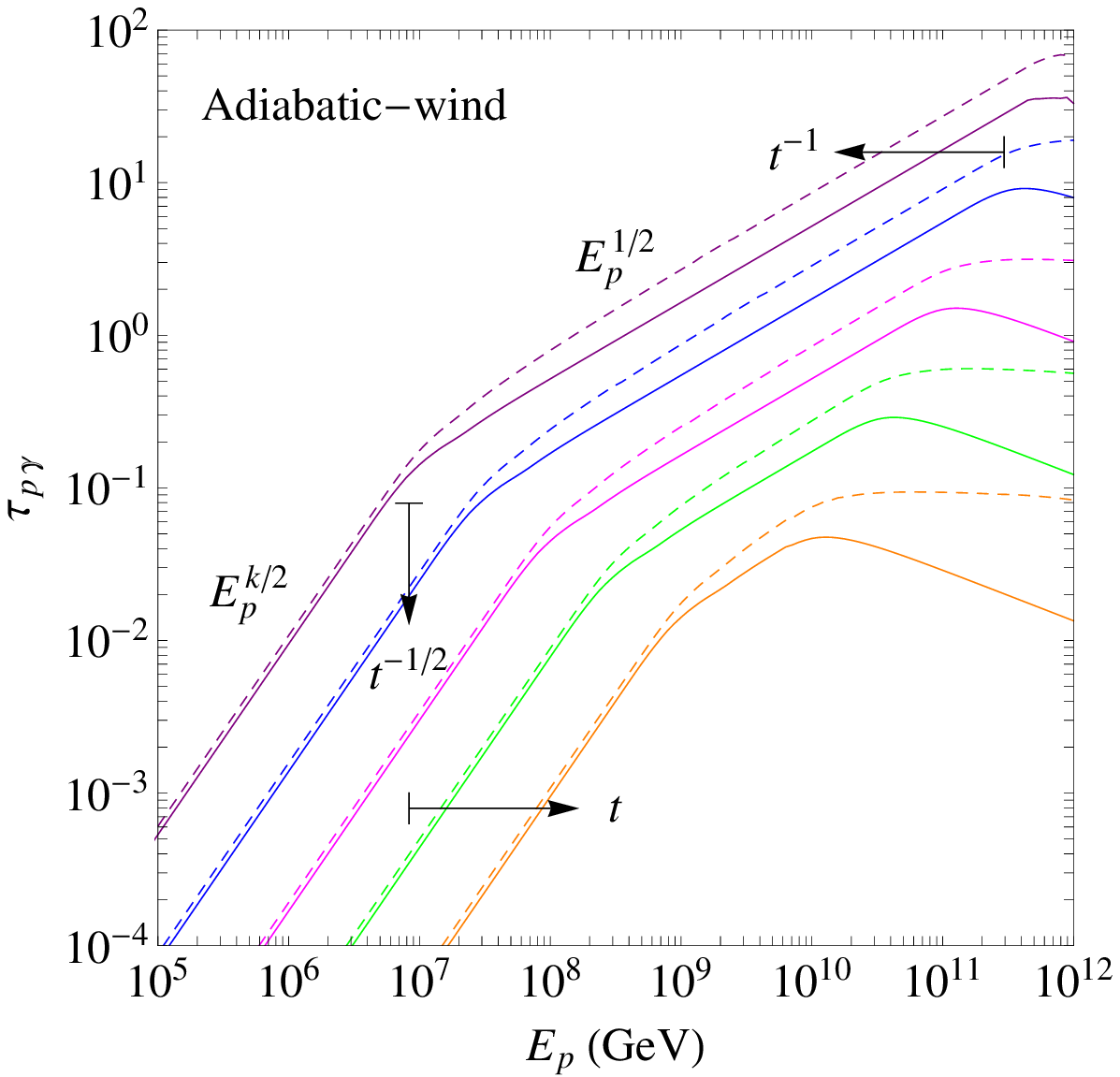}
\includegraphics[width=3.in]{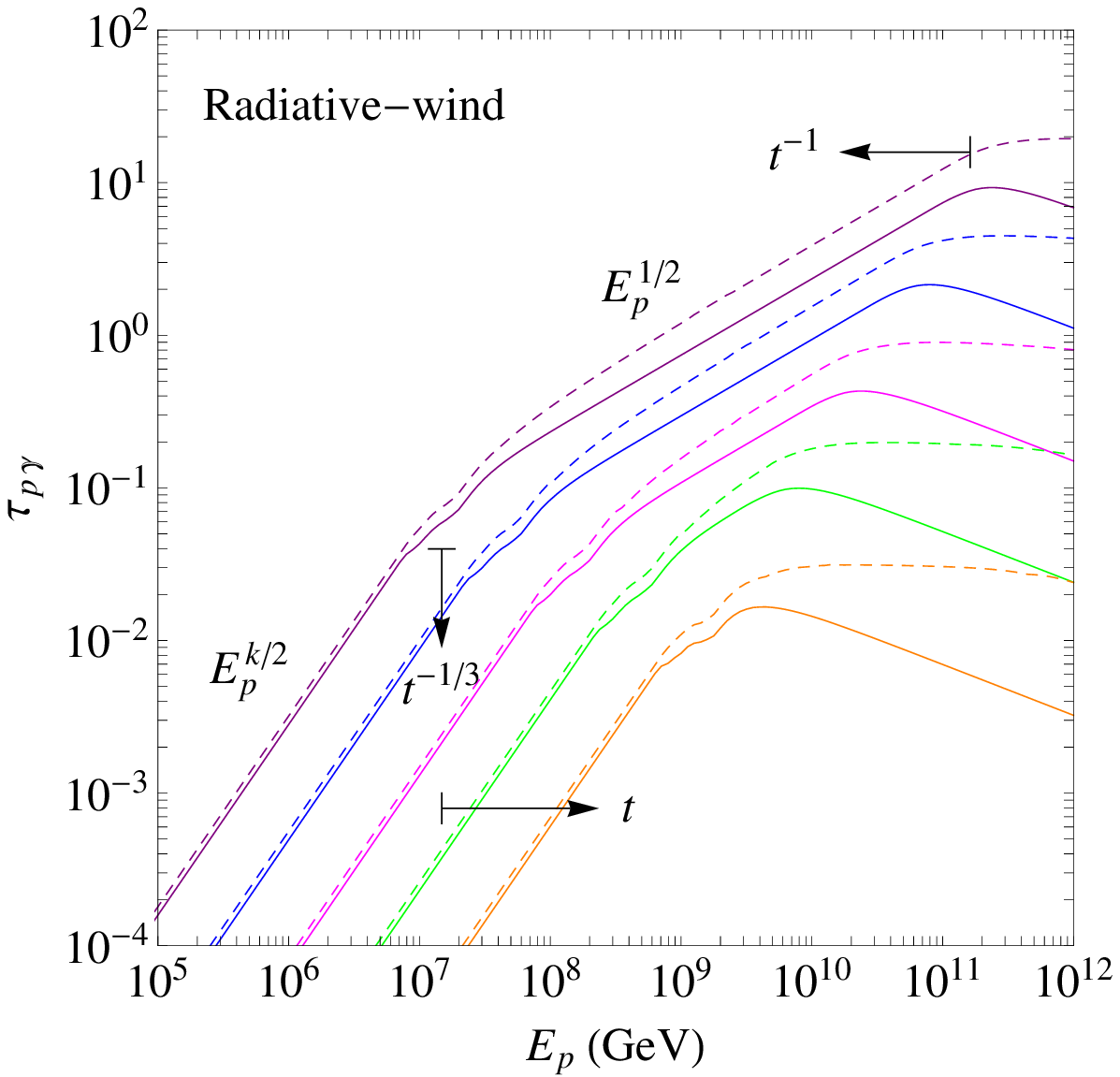}
\caption{Opacity of $p\g$ interactions in the GRB blast wave for
  cosmic-ray protons which are accelerated by the forward shock.  The
  solid and dashed curves correspond to the opacity calculated by the
  $\Delta(1232)$ resonance $p\g$ cross section and by the full $p\g$
  cross section, respectively.  Each panel corresponds to a particular
  blast wave evolution scenario.  The pairs of solid and dashed curves
  are calculated, from left to right, at times $t=t_{dec}$,
  $10t_{dec}$, $10^2t_{dec}$, $10^3t_{dec}$ and $10^4t_{dec}$ for the
  adiabatic cases (upper panels); and at times $t=t_{dec}$,
  $3t_{dec}$, $10t_{dec}$, $30t_{dec}$ and $10^2t_{dec}$ for the
  radiative cases (lower panels).  Temporal behavior of the break
  energies [Eq.~(\ref{proton_breaks})] and $\tau_{p\g}$ at the lower
  break energy [Eq.~(\ref{opt_simple})] are indicated with labeled
  arrows.  Approximate power-law behavior of different segments of the
  curves are also indicated.  Here we have used $k=2.5$ in
  Eq.~(\ref{fast_cooling_spectrum}).}
\label{fig:opacity}
\end{figure}

\end{widetext}

Figure~\ref{fig:opacity} shows the $p\g$ opacity as a function of the
cosmic-ray proton energy, as would be observed if they could escape
and reach us freely, at different times: $t\ge t_{dec}$ for the 4
different blastwave evolution models, adiabatic and radiative blast
waves in the constant density ISM and in the wind of $R^{-2}$ density
profile. For the ISM case we have used $\Gamma_0 = 10^{2.8}$ and
$n_0=1$ whereas in the wind case we have used $\Gamma_0 = 10^{2.6}$
and $A_\star =0.1$.  The other parameters, common to both cases, are
$E_k = 10^{55}$~erg, $\eps_e = \eps_B = 0.1$, $\phi = 10$, $k=2.5$,
$z=1$ and $d_{28} = 2.047$.  The blast wave deceleration time scales
are $10.5$~s and $11.6$~s, respectively, for the ISM and wind cases.
The solid lines are for the $\Delta(1232)$ resonance $p\g$ cross
section and the dashed lines are for the full $p\g$ cross section from
the SOFIA code \cite{Mucke:1999yb}. All curves plotted in each blast
wave model generally have 2 breaks, the lower (higher) energy break
$E_{p,l}$ ($E_{p,h}$) correspnds to the $\veps_m$ ($\veps_c$) in the
synchrotron spectra in Eq.~(\ref{fast_cooling_spectrum}).  A break at
even lower energy corresponding to $\nu_s$ is not shown in the plots.
As noted in Ref.~\cite{Razzaque:2006qa}, given a target photon
spectrum $\veps^{-\alpha}$, the $p\g$ opacity for $\Delta(1232)$
resonance cross section scales as $\propto E_p^{\alpha -1}$.  This
behavior is seen for the 3 power-law segments for opacities with the
$\Delta(1232)$ resonance cross section and below $E_{p,h}$ for the
full cross section.  Contributions by additional channels to
$\sigma_{p\g}$ at $\veps^\p_r > \veps^\p_{r,\rm pk}$ affect
significantly the $\tau_{p\g}$ at $E_p \gtrsim E_{p,l}$ because of
relatively flatter distributions of target photons below $\veps_m$.
For $E_p \lesssim E_{p,l}$, the $\Delta(1232)$ resonance cross section
is still a good approximation for $\tau_{p\g}$ (the difference with
the full cross section is $\lesssim 10\%$) because of a steeply
falling photon spectrum above $\veps_m$.

Here we comment in more details about the temporal behavior of the
break energies and the opacities as indicated by the arrows in
Fig.~\ref{fig:opacity}.  The break energies can be approximately
calculated from the pion production at the energy $\veps^\p_{r,\rm pk}
= 0.3$ GeV of the peak of the $\Delta(1232)$ cross section, from the
condition $\veps^\p_r = 2\g^\p_p \veps^\p = \veps^\p_{r,\rm pk}$ as
\be
E_{p,l/h} \approx \frac{\veps^\p_{r,\rm pk} \Gamma^2}
{2\veps_{m/c} (1+z)^2}.
\label{proton_breaks}
\ee
These break energies are given in Eqs.~(\ref{CRbreaks_ad_i}),
(\ref{CRbreaks_ra_i}), (\ref{CRbreaks_ad_w}) and (\ref{CRbreaks_ra_w})
for different blast wave models and for the reference parameters.  An
approximate analytic expression for the $\tau_{p\g}$ optical depth at
$E_{p,l}$ can be written as
\be
\tau_{p\g}(E_{p,l}) = 
\frac{n^\p_\g (\veps^\p_m) \veps^\p_m \sigma_{\rm pk}R}
{2a\Gamma}.
\label{opt_simple}
\ee
These optical depths are also given in Eqs.~(\ref{opt_ad_i}),
(\ref{opt_ra_i}), (\ref{opt_ad_w}) and (\ref{opt_ra_w}) for the 4
different blast wave models that we consider.  The agreements between
the analytic expressions and the numerical results are very good.

\section{Neutrino flux calculation}

Neutrino flux on the Earth from the GRB blastwave depends on the
efficiency of the $p\gamma$ process, as we have discussed above, and
on the cosmic-ray density (mostly protons) in the blastwave which we
discuss next.

\subsection{Cosmic rays in GRB blastwave}

The total energy of the cosmic rays in the blastwave, after
deceleration ($t>t_{dec}$), is given by
\ba
{\cal E}_{CR}  = \begin{cases} 
\frac{4}{3}\pi \eps_p n_0 R^3(t) m_pc^2 [\Gamma^2 (t) -1] & {\rm ISM} \cr
4\pi \eps_p AR(t) m_pc^2 [\Gamma^2 (t) -1] & {\rm wind},
\end{cases}
\label{total_CR}
\ea
in case of a constant density ISM and $R^{-2}$ wind environment,
respectively.  Here $\eps_p$ is the fraction of blastwave kinetic
energy that goes into accelerated protons.  In case of an adiabatic
blastwave, either in the ISM or in the wind environment, ${\cal
  E}_{CR}\approx \eps_p E_{k}/2$, is constant for $\G(t)\gg1$.  In
case of a radiative blastwave, however, ${\cal E}_{CR}$ evolves with
time as given in Eqs.~(\ref{totCR_ra_i}) and (\ref{totCR_ra_w}),
respectively in the ISM and in the wind environment, respectively.

The energy density of cosmic ray protons in the blastwave is therefore
$u_p = {\cal E}_{CR}/V$ in the local rest frame, where $V=(4/3)\pi
R^3$ is the volume.  Acceleration of protons in the GRB blastwave
to ultrahigh energies has been discussed in the past
\cite{Vietri:1995hs}.  Here we assume that the differential number
density of protons, with $n(E_p)\propto E_p^{-2}$ spectrum expected
from shock acceleration, is
\be
n(E_p) = \frac{{\cal E}_{CR}}
{V E_p^2 {\rm ln}(\g^\p_{p,s}/\g^\p_{p,m})}. 
\label{CR_spectrum}
\ee
Here $\g^\p_{p,m} = \Gamma$ is the minimum proton Lorentz factor and
$\g^\p_{p,s}$ is the saturation proton Lorentz factor, both in the
comoving blastwave frame.  We derive $\g^\p_{p,s}$, from the condition
that the proton acceleration time $t^\p _{acc} = \phi \g^\p_p
m_pc/(eB^\p)$ is limited by the dynamic time $t^\p_{dyn} =
t\Gamma/(1+z)$, as
\be
\g^\p_{p,s} (t) = \frac{eB^\p (t)}{\phi m_pc} \frac{t\Gamma(t)}{1+z}.
\label{saturation_p}
\ee
Here $\phi$ is the number of gyroradius required to accelerate proton
to the saturation Lorentz factor.  An observer would measure an energy
$E_{p,s} = \g^\p_{p,s} \Gamma/(1+z)$, if these protons could escape
the acceleration cite as cosmic rays to reach us, and are given in
Eqs.~(\ref{Emax_ad_i}), (\ref{Emax_ra_i}), (\ref{Emax_ad_w}) and
(\ref{Emax_ra_w}) for the 4 different scenarios that we consider.
Note that, in case of strong magnetic field in the blastwave,
$t^\p_{acc}$ could be limited by the synchrotron cooling time of the
proton $t^\p_{p,syn} = (m_p/m_e)^3 (6\pi m_e c)/(\sigma_T B^{\p 2}
\g^\p_p)$, rather than $t^\p_{dyn}$.  In such a case the proton
saturation Lorentz factor would be given by $\g^\p_{p,s} = (m_p/m_e)
\sqrt{6\pi e/(\phi \sigma_T B^\p)}$.

If the cosmic-ray protons could escape freely from the blastwave and
avoid interactions with CMB photons, their flux on the Earth would be
\be
J_p (E_p) = \frac{c}{4\pi} \left( \frac{R}{d_L} \right)^2 n(E_p),
\label{CR_flux}
\ee
in the zero galactic and intergalactic magnetic field.  This flux for
the 4 different blaswave scenarios are given in Eq.~(\ref{Jp_ad_i}),
(\ref{Jp_ra_i}), (\ref{Jp_ad_w}) and (\ref{Jp_ra_w}) with the
logarithmic factor in Eq.~(\ref{CR_spectrum}) given by $\xi_{1} = {\rm
  ln}(\g^\p_{p,s}/\g^\p_{p,m})/10$.  Note that all these fluxes
decrease with time.

\subsection{Neutrino fluxes on the Earth}

UHE $\nu$'s from the $p\g \to n\pi^+$ interactions are produced in two
steps, first via $\pi^+ \to \mu^+ \nu_\mu \to e^+ \nu_e {\bar \nu}_\mu
\nu_\mu$ chain decay in the blastwave, and second via neutron beta
decay process $n \to pe^- {\bar \nu}_e$ by escaping neutrons from the
blastwave while on their way to the Earth. We ignore a small
contribution by the $n$-decay flux component in our calculation for
simplicity.  This component, however, could be important for neutrino
flavor ratio calculations.  Furthermore we calculate neutrino fluxes
from the $\Delta (1232)$ resonance $p\g$ cross section, as a
conservative estimate.  Calculations with the full cross section gives
a $\lesssim 30\%$ higher flux in the PeV--EeV range of our interest.

The flux of secondary $\pi^+$ or $\pi^0$ can be calculated in general,
at a time $t\gtrsim t_{dec}$, as
\ba 
J_\pi (E_\pi) &=& \int_0^1 \frac{dx}{x} f_{p\to \pi} (x)
J_p \left(\frac{E_\pi}{x}\right) \nonumber \\
&& \times K_{p\g} \left(\frac{E_\pi (1+z)}{x\Gamma}\right)
\frac{t\Gamma}{1+z} ;~ x=\frac{E_\pi}{E_p}.
\label{pion_yeld}
\ea
Here $K_{p\g}$ is evaulated in the blastwave frame.  A similar
expression can be used for the secondary neutron flux by replacing $\pi
\to n$ in the above equation.  A simple expression of
Eq.~(\ref{pion_yeld}) follows from the assumption of the pion yield
function $f_{p\to\pi} = \delta (x-\langle x\rangle)/2$ for an equal
probability of $\pi^+$ and $\pi^0$ production with a mean inelasticity
$\langle x\rangle \approx 0.2$.  Therefore, using
Eq.~(\ref{pg_opacity}), we get
\be
J_\pi (E_\pi) \approx \frac{1}{2\langle x\rangle} 
J_p \left(\frac{E_\pi}{\langle x\rangle}\right)
\tau_{p\g} \left(\frac{E_\pi (1+z)}{\langle x\rangle\Gamma}\right),
\label{pi_flux}
\ee
for $\tau_{p\g} \le 1$.  For the neutron flux, $J_n (E_n)$, the mean
inelasticity is $\langle x\rangle \approx 0.8$.

\begin{widetext}

\begin{figure}[t]
\includegraphics[width=3.in]{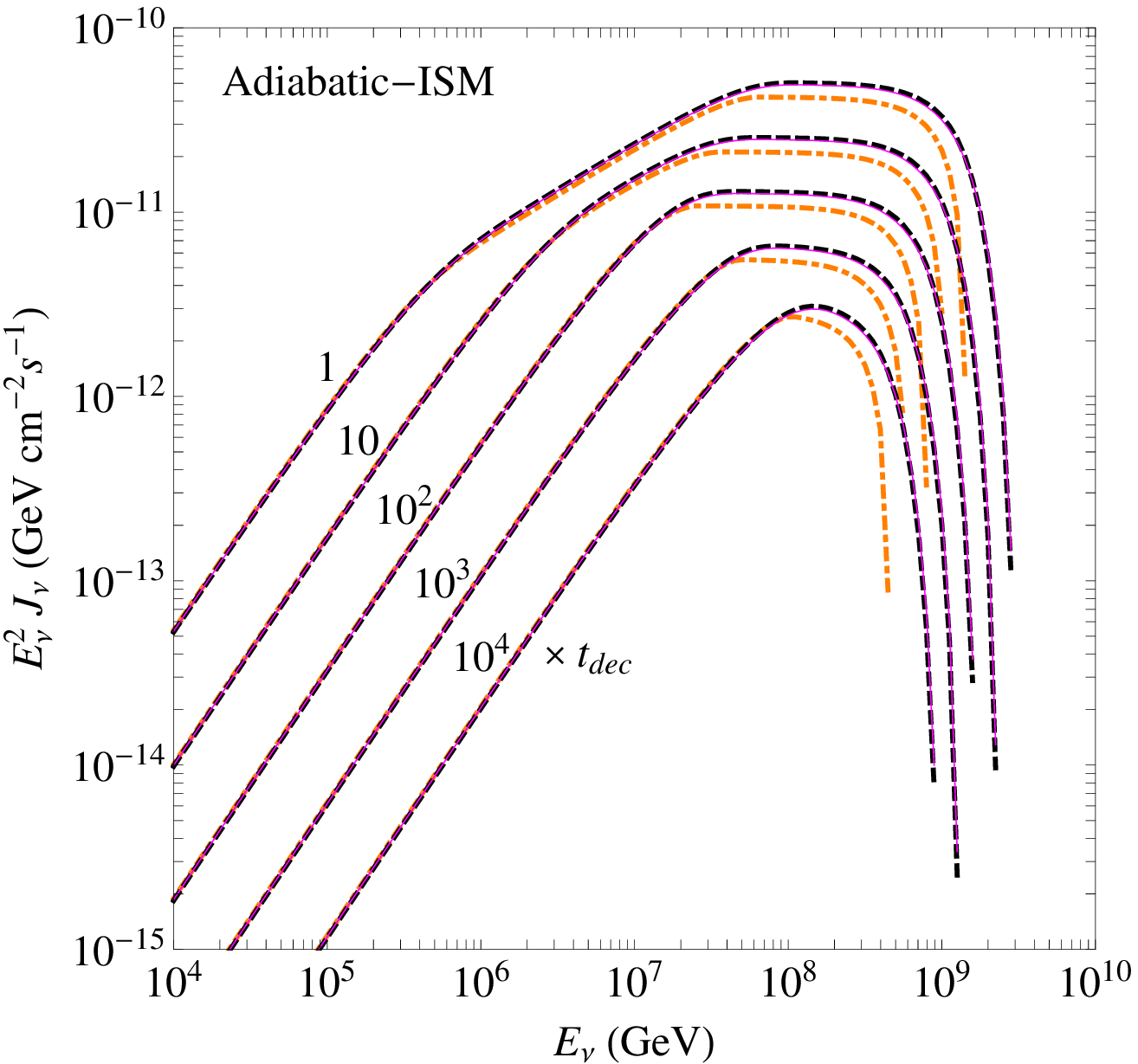}
\includegraphics[width=3.in]{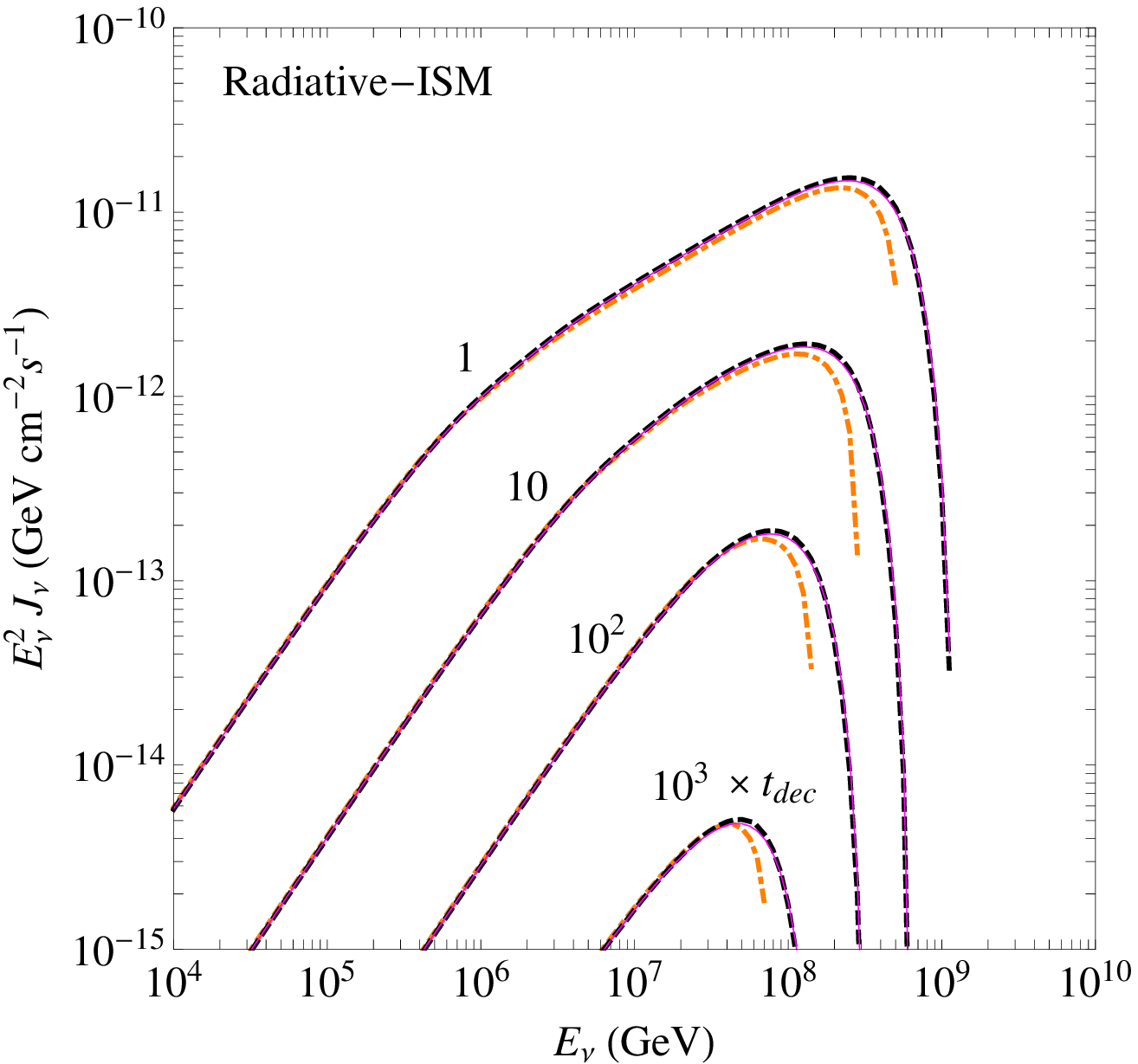}
\includegraphics[width=3.in]{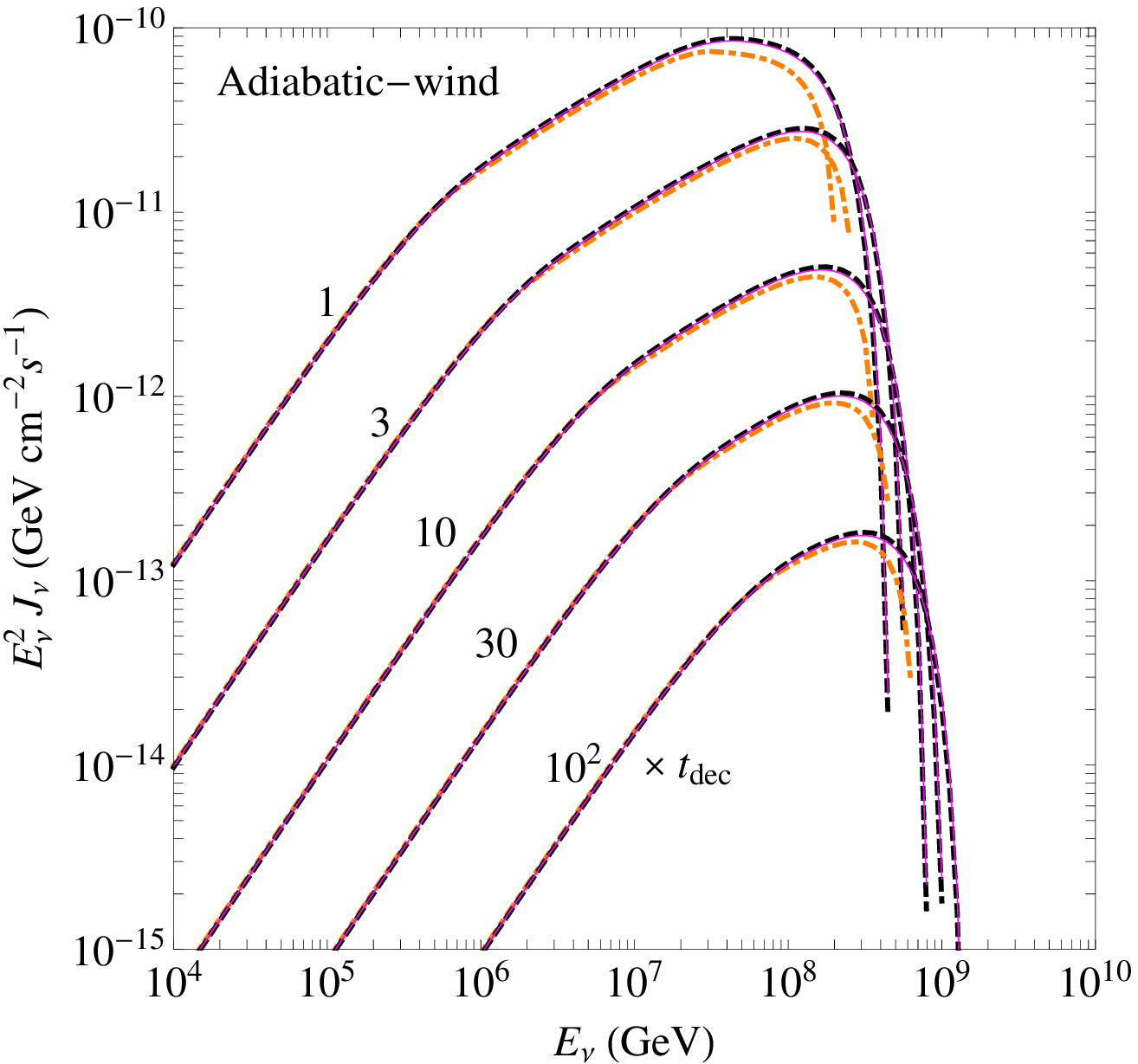}
\includegraphics[width=3.in]{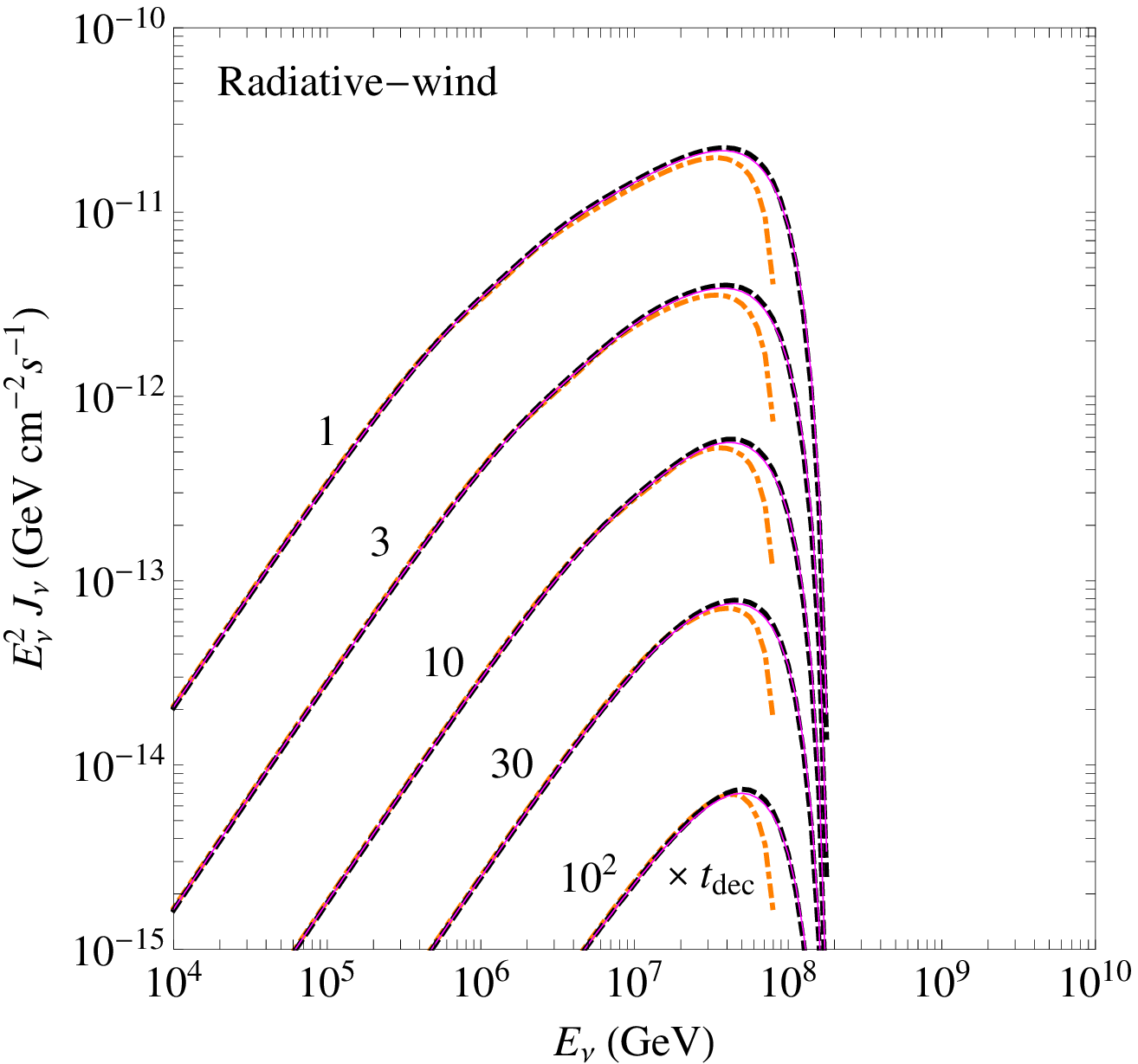}
\caption{Neutrino fluxes from a GRB at redshift $z=1$ for 4 different
  blastwave models.  The model parameters are the same as in
  Fig.~\ref{fig:opacity} for each case: $\Gamma_0 = 10^{2.8}$ and $n_0
  = 1$ for the ISM ($t_{dec} = 10.5$~s); $\Gamma_0 = 10^{2.6}$ and
  $A_\star = 0.1$ for the wind ($t_{dec} = 11.6$~s); the common
  parameters are $E_k = 10^{55}$~erg, $\eps_p = 1$, $\eps_e = \eps_B =
  0.1$, $\phi = 10$, $k=2.5$ and $d_{L} = 2.047\times 10^{28}$~cm.
  The dotted-dash lines are for $\nu_\mu$ fluxes from $\pi^+$ decay,
  the solid lines are for ${\bar \nu}_\mu$ fluxes from $\mu^+$ decay
  and the dashed lines are for $\nu_e$ fluxes from $\mu^+$ decay.
  Neutrino oscillation has not been taken into account for the plotted
  fluxes.}
\label{fig:nuflux}
\end{figure}

\end{widetext}

The muon and muon neutrino (pionic neutrino) fluxes from the pion
decay $\pi^+ \to \mu^+  \nu_\mu$ are given by
\ba
J_\mu (E_\mu) &=& \int_0^1 \frac{dx}{x} f_{\pi^+ \to \mu^+} (x) 
J_\pi \left( \frac{E_\mu}{x} \right) ;~ x=\frac{E_\mu}{E_\pi}, 
\nonumber \\
J_{\nu_\mu} (E_\nu) &=& \int_0^1 \frac{dx}{x} f_{\pi^+ \to \nu_\mu} (x)
J_\pi \left(\frac{E_\nu}{x} \right) ;~
x=\frac{E_\nu}{E_\pi}. ~~
\label{pion_decay_fluxes}
\ea
where the scaling functions $f_{\pi^+ \to \mu^+}$ and $f_{\pi^+ \to
  \nu_\mu}$ are given by Eq.~(\ref{pi_mu_dist}), following
Ref.~\cite{Lipari:1993hd}.  The subsequent $\nu$ fluxes (muonic
neutrinos) from the $\mu^+ \to e^+ \nu_e {\bar \nu}_\mu$ decay are
given by
\ba
J_{\nu_e} (E_\nu) &=&
\int_0^1 \frac{dy}{y} \int_0^1 \frac{dx}{x}
f_{\mu^+ \to \nu_e} (x,y) f_{\pi \to \mu} (x)
\nonumber \\ &&\times
J_\pi \left( \frac{E_\nu}{xy} \right) ;~
x=\frac{E_\mu}{E_\pi},~ y=\frac{E_\nu}{E_\mu},
\nonumber \\
J_{{\bar \nu}_\mu} (E_\nu) &=& 
\int_0^1 \frac{dy}{y} \int_0^1 \frac{dx}{x} 
f_{\mu^+ \to {\bar \nu}_\mu} (x,y) f_{\pi \to \mu} (x)
\nonumber \\ &&\times
J_\pi \left( \frac{E_\nu}{xy} \right) ;~
x=\frac{E_\mu}{E_\pi},~ y=\frac{E_\nu}{E_\mu},
\label{muon_decay_fluxes} 
\ea 
The scaling functions $f_{\mu^+ \to \nu_e}$ and $f_{\mu^+ \to {\bar
    \nu}_\mu}$ from Ref.~\cite{Lipari:1993hd} are given in
Eq.~(\ref{mu_nu_dist}) for completeness.

Figure \ref{fig:nuflux} shows $\nu$ fluxes from a GRB at redshift
$z=1$ for the 4 different blastwave evolution models that we have
considered.  The fluxes are calculated at the same time as for the
respective $\tau_{p\g}$ plots in Fig.~\ref{fig:opacity}.  The fluxes
for the adiabatic blastwave in the wind environment at $t_{dec} =
11.6$~s are the highest among all 4 blastwave models, in the PeV--EeV
range.  On the other hand, significantly high fluxes from an adiabatic
blastwave in the ISM environment last for the longest time.  Fluxes
from a radiative blastwave, either in the ISM or wind environment,
decrease faster than the fluxes from an adiabatic fireball, as the
$p\g$ opacity also dcereases faster in the radiative blastwaves
(Fig.~\ref{fig:opacity}).  Plotted fluxes in Fig.~\ref{fig:nuflux} are
the ``source fluxes'' without taking into account neutrino
oscillation.  To a good approximation, the $\nu_e + {\bar \nu}_e$,
$\nu_\mu + {\bar \nu}_\mu$ and $\nu_\tau + {\bar \nu}_\tau$ fluxes at
a detector on the Earth will be equal to the plotted ${\bar \nu}_\mu$
flux in Fig.~\ref{fig:nuflux}.

It is not straightforward to estimate the diffuse fluxes of neutrinos
from the GRB blastwave models we have considered, because of the
unknown rate of these bursts.  Our models are motivated by the {\it
Fermi}-LAT detection of delayed emission of GeV photons from GRBs.
Such delays are explained as forward-shock synchrotron emission
\cite{Kumar:2009vx, Ghisellini+10, Razzaque:2010ku}, requiring $\sim
10$~s time scale for long-duration GRB fireball with high bulk Lorentz
factor to decelerate.  Detection of GRBs by {\it Fermi}-LAT during its
operation since launch in 2008 suggests that the rate of the GeV
bright GRBs is likely lower than the rate of typical GRBs
\cite{Fermi-LAT:2013kxa}.

\begin{figure}[t]
\includegraphics[width=3.in]{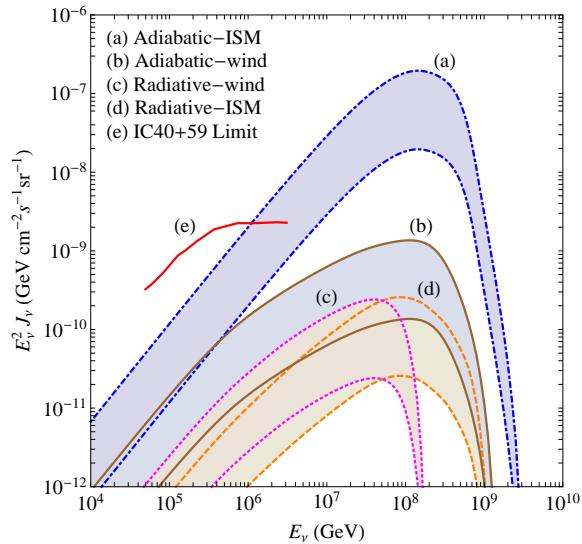}
\caption{Diffuse $\nu_\mu + {\bar \nu}_\mu$ fluxes from GRB blastwave
  in 4 different evolution scenarios.  The top and bottom lines of the
  same type, with shaded region in between, correspond to $100\%$ and
  $10\%$ of the GRBs having high bulk Lorentz factors. Also shown is
  the recent upper limit on GRB internal shock $\nu$ flux model
  \cite{Waxman:1997ti} from IceCube \cite{Abbasi:2012zw}.  Neutrino
  oscillation in vacuum has been taken into account for the diffuse
  flux models, which leads to equal fluxes of $\nu_e + {\bar \nu}_e$
  and $\nu_\tau + {\bar \nu}_\tau$ as the $\nu_\mu + {\bar \nu}_\mu$
  flux plotted here.}
\label{fig:diffflux}
\end{figure}

In Figure \ref{fig:diffflux} we roughly estimate the diffuse $\nu$
fluxes as follows. We calculate the time-integrated flux from the
fluxes plotted in Fig.~\ref{fig:nuflux} for each of the blastwave
models.  We assume the rate of these GRBs, placed at redshift $z=1$,
is 2/day over the whole sky and cosmological source evolution gives a
multiplicative factor of 3.  To show the uncertainty of the rate of
GRBs with high bulk Lorentz factor, we have plotted the diffuse fluxes
assuming that $100\%$ and $10\%$ of the GRBs have the same
characteristics used in modeling. These are indicatives of the rates
detected by {\em Fermi}-GBM and {\em Fermi}-LAT, respectively. The
plotted $\nu_\mu + {\bar \nu}_\mu$ fluxes are calculated by taking
into account neutrino oscillation in vacuum, which gives nearly equal
fluxes of the 3 flavors.  Note that the diffuse $\nu$ flux from the
adibatic blastwave in ISM dominates the flux models, as expected from
the longest-lived emission in this scenario.  We have also shown the
recently published IceCube upper limit on the GRB prompt $\nu$ flux
\cite{Abbasi:2012zw}.  Notably our $\nu$ flux models are consistent
with this limit, except for the adiabatic blastwave in the ISM case at
$\sim 1$--3 PeV.  This might have interesting consequences with
regards to the recent discovery of two neutrino events at $\sim 1$ PeV
\cite{Aartsen:2013bka} by the IceCube Neutrino Observatory
\cite{Ahrens:2003ix}.

\section{Detection prospects}

It was pointed out sometime ago that the burst-to-burst fluctuation
may result in single GRBs to dominate the diffuse flux
\cite{AlvarezMuniz:2000st}.  Moreover, only a few neutrinos might be
detected by IceCube, in the optimistic Waxman-Bahcall scenario
\cite{Waxman:1997ti}, for a very nearby burst such as GRB 030329
\cite{Razzaque:2003uw} at redshift $z=0.17$.  It is unlikely that the
two $\sim 1$ PeV events detected by IceCube originated from very
nearby GRBs, using the flux models presented here and given the
average neutrino effective area ($\sim 10$ m$^2$ at $\sim 1$ PeV for
the combined 3 flavors) of the detector when these two events were
detected \cite{Aartsen:2013bka}.

The combined diffuse flux of all 3 $\nu$ flavors in case of the
adiabatic blastwave in ISM (Fig.~\ref{fig:diffflux}) is $\sim 6\times
10^{-9}$~GeV~cm$^{-2}$~s$^{-1}$~sr$^{-1}$ at 1 PeV.  A $\sim 6$ times
higher cosmological GRB rate than the 2/day that we have used or a
$\sim 6$ times higher kinetic energy per burst could in principle
produce the $3.6\times 10^{-8}$~GeV~cm$^{-2}$~s$^{-1}$~sr$^{-1}$
diffuse neutrino flux model required by IceCube to generate the two
$\sim 1$ PeV events \cite{Aartsen:2013bka}.  Note that this required
diffuse flux is higher than the Waxman-Bahcall limit
\cite{Waxman:1998yy}.

The prosepects for detection of PeV--EeV neutrinos from GRB blastwaves
that we have modeled, are much better for the future 100 km$^3$ radio
Askaryan detectors in Antarctica such as ARIANNA \cite{Barwick:2006tg}
and ARA \cite{Allison:2011wk}.  A few PeV--EeV neutrinos can be
detected by these experiments from nearby GRBs according to our flux
models.  Our diffuse $\nu$ flux for the adiabatic-ISM scenario should
be detectable by ARIANNA or ARA within 3 years of their operation.

\section{Discussion and conclusions}

We have calculated new, realistic $\nu$ fluxes from GRB blastwave
models, which are responsible for radio to X-ray afterglow and
possibly GeV $\g$-ray emission.  A very-high bulk Lorentz factor of
the GRB jet, which adversely affects the $\nu$ production efficiency
in the internal shocks, works favorably in our scenario by shortening
the afterglow onset time and by producing a bright afterglow which
provide ample target photons for $p\g$ interactions.  The long-lived
PeV--EeV $\nu$ emission is due to $p\g$ interactions of protons,
accelerated to UHE in the forward shock of the blastwave, with
afterglow photons.  These $\nu$ fluxes from the external forward shock
should be present in the GRB afterglow, if protons are co-accelerated
with electrons, even if the prompt $\g$ rays are produced by a
different mechanism than the internal shocks and/or if the external
reverse shock is absent, e.g.\ in case of a Poynting flux dominated
GRB ejecta.

We have computed $\nu$ fluxes for 4 different blastwave evolution
scenarios, namely adiabatic and fully radiative blastwaves in a
constant density ISM environment and in an environment with $R^{-2}$
wind density profile.  The PeV--EeV $\nu$ fluxes peak at the blastwave
deceleration time and decrease after that in all the cases, as the
$p\g$ interaction efficiency decreases with the increasing blastwave
radius, and as the protons are less efficiently accelerated to UHE
with increasing time.  Neutrino fluxes from the adiabatic blastwaves
last longer than the radiative ones, as expected, with the
adiabatic-ISM scenario being the longest lasting case.  The diffuse
$\nu$ fluxes that we have calculated depend on the unknown rate of the
high bulk Lorentz factor bursts.  The diffuse $\nu$ flux from an
adiabatic blastwave in ISM is the highest among all the models.

The interpretation of the two $\sim 1$ PeV $\nu$ events detected by
IceCube does not follow naturally from our diffuse flux models but
could be accommodated in very optimistic scenario with higher GRB rate
and/or higher kinetic energy per GRB than we have considered.
Detection of the PeV--EeV $\nu$ from the GRB afterglow that we have
predicted could be possible by upcoming, large radio Askaryan
detectors in Antarctica and verify the hypothesis of UHECR
acceleration in the GRB blastwave.

\acknowledgments

I would like to thank David Z.\ Besson for useful communication about
the ARA and ARIANNA experiments.  I would also like to thank Zhuo Li,
Kohta Murase and Peter Veres for comments.

\appendix

\begin{widetext}

\section{Blastwave models and synchrotron spectra}
\label{appendA}

In the following, we provide the bulk Lorentz factor and radius of the
blastwave along with the shock magnetic field in four different
evolution scenarios described in Eqs.~(\ref{ism_bulk_Lorentz}) and
(\ref{wind_bulk_Lorentz}).  We also list the break frequencies
($h\nu_c$, $h\nu_m$, $h\nu_s$) in the synchrotron spectrum, the time
scale ($t_0$) for which the fast-cooling regime is valid and the
maximum synchrotron flux ($F_{\nu,m}$), to facilitate calculation of
the target photon spectrum for $p\gamma$ interactions in the GRB
blastwave frame.  We have assumed the parameters $\eps_B =
0.1\eps_{B,-1}$, $\eps_e = 0.1\eps_{e,-1}$ from typical GRB afterglow
modeling, $10\phi_1$ gyroradius to accelerate particles and a
reference time $t=10^2 t_{2}$~s.

\subsection{Adiabatic blastwave in ISM}

\be
\Gamma = 124\,(1+z)^{3/8} n_0^{-1/8} E_{55}^{1/8} t_{2}^{-3/8}.
\label{bw_G_ad_i}
\ee
\be
R = 3.7\times 10^{17}
(1+z)^{-1/4} n_0^{-1/4} E_{55}^{1/4} t_{2}^{1/4} ~{\rm cm}.
\label{bw_R_ad_i}
\ee
\be
B^\p = 15.3\,(1+z)^{3/8} \eps_{B,-1}^{1/2} n_0^{3/8} 
E_{55}^{1/8} t_{2}^{-3/8} ~{\rm G}.
\label{bw_B_ad_i}
\ee
\ba
h\nu_c &=& 2.3\, (1+z)^{-1/2} \eps_{B,-1}^{-3/2} n_0^{-1} 
E_{55}^{-1/2} t_2^{-1/2} ~{\rm eV}, \nonumber \\
h\nu_m &=& 17.3\, (1+z)^{1/2} \eps_{B,-1}^{1/2} \eps_{e,-1}^2 
E_{55}^{1/2} t_2^{-3/2} ~{\rm keV}, \nonumber \\
h\nu_s &=& 2.9\, (1+z)^{-5/8} \phi_1^{-1} n_0^{-1/8} 
E_{55}^{1/8} t_2^{-3/8} ~{\rm GeV}.
\label{break_nu_ad_ism}
\ea
\be
t_{0} = 1.1\times 10^7 (1+z) \eps_{B,-1}^2 \eps_{e,-1}^2 n_0 
E_{55} ~{\rm s}.
\label{t0_ad_ism}
\ee
\ba
F_{\nu,m} = 8.2\, (1+z)^{-1} \eps_{B,-1}^{1/2} n_0^{1/2} 
E_{55} d_{28}^{-2}~{\rm Jy}.
\label{Fnumax_ad_ism}
\ea
%

\subsection{Radiative blastwave in ISM}

\be
\Gamma 
= 86\,(1+z)^{3/7} n_0^{-1/7} \Gamma_{2.5}^{-1/7} E_{55}^{1/7} t_2^{-3/7}.
\label{bw_G_ra_i}
\ee
\be
R = 3.1\times 10^{17}
(1+z)^{-1/7} n_0^{-2/7} \Gamma_{2.5}^{-2/7} E_{55}^{2/7} t_2^{1/7}~{\rm cm}.
\label{bw_R_ra_i}
\ee
\be
B^\p = 10.5\,(1+z)^{3/7} \eps_{B,-1}^{1/2} n_0^{5/14} \Gamma_{2.5}^{-1/7} 
E_{55}^{1/7} t_2^{-3/7} ~{\rm G}.
\label{bw_B_ra_i}
\ee
\ba
h\nu_c &=& 10.4\, (1+z)^{-5/7} \eps_{B,-1}^{-3/2} n_0^{-13/14}
\Gamma_{2.5}^{4/7} E_{55}^{-4/7} t_2^{-2/7} ~{\rm eV},
\nonumber \\ 
h\nu_m &=& 4.0\, (1+z)^{5/7} \eps_{B,-1}^{1/2} \eps_{e,-1}^2 n_0^{-1/14}
\Gamma_{2.5}^{-4/7} E_{55}^{4/7} t_2^{-12/7} ~{\rm keV},
\nonumber \\ 
h\nu_s &=& 2.0\,(1+z)^{-4/7} \phi_1^{-1} n_0^{-1/7} \Gamma_{2.5}^{-1/7} 
E_{55}^{1/7} t_2^{-3/7} ~{\rm GeV}.
\label{break_nu_ra_ism}
\ea
\ba
t_0 &=& 6.3\times 10^3 (1+z) \eps_{B,-1}^{7/5} \eps_{e,-1}^{7/5} n_0^{3/5} 
\Gamma_{2.5}^{-4/5} E_{55}^{4/5} ~{\rm s}.
\ea 
\label{t0_ra_ism}
\ba
F_{\nu,m} = 2.2\, (1+z)^{-4/7} \eps_{B,-1}^{1/2} n_0^{5/14} 
\Gamma_{2.5}^{-8/7} E_{55}^{8/7} t_{2}^{-3/7} d_{28}^{-2} ~{\rm Jy}.
\label{Fnumax_ra_ism}
\ea
%

\subsection{Adiabatic blastwave in wind}

\be
\Gamma = 78\,(1+z)^{1/4} A_\star^{-1/4} E_{55}^{1/4} t_2^{-1/4}.
\label{bw_G_ad_w}
\ee
\be
R = 1.4\times 10^{17} 
(1+z)^{-1/2} A_\star^{-1/2} E_{55}^{1/2} t_2^{1/2}~{\rm cm}.
\label{bw_R_ad_w}
\ee
\be
B^\p =  9.5\,(1+z)^{1/4} \eps_{B,-1}^{1/2} A_\star^{1/4} 
E_{55}^{1/4} t_2^{-1/4} ~{\rm G}.
\label{bw_B_ad_w}
\ee
\ba
h\nu_c &=& 0.3\, (1+z)^{-3/2} \eps_{B,-1}^{-3/2} A_\star^{-2} 
E_{55}^{1/2}t_2^{1/2} ~{\rm eV}, 
\nonumber \\
h\nu_m &=& 10.0\,(1+z)^{1/2} \eps_{B,-1}^{1/2} \eps_{e,-1}^2 
E_{55}^{1/2} t_2^{-3/2} ~{\rm keV},
\nonumber \\
h\nu_s &=& 1.8\, (1+z)^{-3/4} \phi_1^{-1} A_\star^{-1/4} 
E_{55}^{1/4} t_2^{-1/4} ~{\rm GeV}.
\label{break_nu_ad_wind}
\ea
\be
t_0 = 1.9\times 10^4 (1+z) \eps_{B,-1} \eps_{e,-1} A_\star ~{\rm s}.
\label{t0_adiabatic_wind}
\ee
\ba
F_{\nu_m} &=& 10.4\, (1+z)^{-1/2} \eps_{B,-1}^{1/2} A_\star 
E_{55}^{1/2} t_2^{-1/2} d_{28}^{-2}  ~{\rm Jy}.
\label{Fnumax_ad_wind}
\ea
%

\subsection{Radiative blastwave in wind}

\be
\Gamma = 
40\,(1+z)^{1/3} A_\star^{-1/3} \Gamma_{2.5}^{-1/3} E_{55}^{1/3} t_2^{-1/3}.
\label{bw_G_ra_w}
\ee
\be
R = 6.9\times 10^{16}
(1+z)^{-1/3} A_\star^{-2/3} \Gamma_{2.5}^{-2/3} E_{55}^{2/3} t_2^{2/3}
~{\rm cm}.
\label{bw_R_ra_w}
\ee
\be
B^\p = 5.0\,(1+z)^{1/3} \eps_{B,-1}^{1/2} A_\star^{1/6} 
\Gamma_{2.5}^{-1/3} E_{55}^{1/3} t_2^{-1/3} ~{\rm G}.
\label{bw_B_ra_w}
\ee
\ba 
h\nu_c &=& 0.4\,(1+z)^{-4/3} \eps_{B,-1}^{-3/2} A_\star^{-13/6} 
\Gamma_{2.5}^{-2/3} E_{55}^{2/3} t_2^{1/3} ~{\rm eV}, 
\nonumber \\ 
h\nu_m &=& 1.5\, (1+z)^{2/3} \eps_{B,-1}^{1/2} \eps_{e,-1}^2
A_\star^{-1/6} \Gamma_{2.5}^{-2/3} E_{55}^{2/3} t_2^{-5/3} ~{\rm keV},
\nonumber \\ 
h\nu_s &=& 1.0\, (1+z)^{-2/3} \phi_1^{-1} A_\star^{-1/3} 
\Gamma_{2.5}^{-1/3} E_{55}^{1/3} t_2^{-1/3} ~{\rm GeV}.
\label{break_nu_ra_wind}
\ea
\be
t_0 =  6.1\times 10^3 (1+z) \eps_{B,-1} \eps_{e,-1} A_\star \,{\rm s}.
\label{t0_radiative_wind}
\ee
\ba
F_{\nu_m} &=& 2.8\, (1+z)^{-1/3} \eps_{B,-1}^{1/2} A_\star^{5/6} 
\Gamma_{2.5}^{-2/3} E_{55}^{2/3} t_2^{-2/3} d_{28}^{-2}  ~{\rm Jy}.
\label{Fnumax_ra_wind}
\ea
%

\section{$p\gamma$ interaction and cosmic-ray parameters}
\label{appendB}

Here we provide numerical values for the break energies in
Eq.~(\ref{proton_breaks}), the optical depth in
Eq.~(\ref{opt_simple}), the total energy in cosmic rays given by
Eq.~(\ref{total_CR}), the limiting cosmic-ray energy in
Eq.~(\ref{saturation_p}) and the cosmic-ray flux in
Eq.~(\ref{CR_flux}) for the 4 different blast wave models.

\subsection{Adiabatic blastwave in ISM}

\ba
E_{p,l} &=&  1.3\times 10^{8} (1+z)^{-7/4} \eps_{B,-1}^{-1/2} 
\eps_{e,-1}^{-2} n_0^{-1/4} E_{55}^{-1/4} t_2^{3/4} ~{\rm GeV},
\nonumber \\
E_{p,h} &=&  1.0\times 10^{12} (1+z)^{-3/4} \eps_{B,-1}^{3/2} 
n_0^{3/4} E_{55}^{3/4} t_2^{-1/4} ~{\rm GeV}.
\label{CRbreaks_ad_i}
\ea
\be
\tau_{p\g} (E_{p,l}) = 0.7\, (1+z)^{-1/2} \eps_{B,-1}^{1/2} 
n_0 E_{55}^{1/2} t_{2}^{1/2}. 
\label{opt_ad_i}
\ee
\be
{\cal E}_{CR} = \eps_p E_k/2.
\label{totCR_ad_i}
\ee
\be
E_{p,s} = 2.3\times 10^{19} (1+z)^{-7/8} \phi_1^{-1} n_0^{1/8} 
\eps_{B,-1}^{1/2} E_{55}^{3/8} t_2^{-1/8} ~{\rm eV}.
\label{Emax_ad_i}
\ee
\be 
E_p^2 J_p(E_p) = 4.8\times 10^{-9} (1+z)^{1/4} \xi_1^{-1} \eps_p
n_0^{1/4} E_{55}^{3/4} t_2^{-1/4} d_{28}^{-2} ~{\rm GeV~cm}^{-2}~{\rm
  s}^{-1}.
\label{Jp_ad_i}
\ee
%

\subsection{Radiative blastwave in ISM}

\ba
E_{p,l} &=&  2.8\times 10^{8} (1+z)^{-13/7} \eps_{B,-1}^{-1/2} 
\eps_{e,-1}^{-2} n_0^{-3/14} \Gamma_{2.5}^{2/7} E_{55}^{-2/7} 
t_2^{6/7} ~{\rm GeV},
\nonumber \\
E_{p,h} &=&  1.1\times 10^{11} (1+z)^{-3/7} \eps_{B,-1}^{3/2} 
n_0^{9/14} \Gamma_{2.5}^{-6/7} E_{55}^{6/7} t_2^{-4/7} ~{\rm GeV}.
\label{CRbreaks_ra_i}
\ea
\be
\tau_{p\g} (E_{p,l}) = 0.5\, (1+z)^{-2/7} \eps_{B,-1}^{1/2}
n_0^{13/14} \Gamma_{2.5}^{-4/7} E_{55}^{4/7} t_2^{2/7}.
\label{opt_ra_i}
\ee
\be
{\cal E}_{CR} = 1.4\times 10^{54} (1+z)^{3/7} \eps_p n_0^{-1/7}  
\Gamma_{2.5}^{-8/7} E_{55}^{8/7} t_{2}^{-3/7}  ~{\rm erg}.
\label{totCR_ra_i}
\ee
\be 
E_{p,s} = 7.4\times 10^{18} (1+z)^{-5/7} \phi_1^{-1}
\eps_{B,-1}^{1/2} n_0^{1/14} \Gamma_{2.5}^{-3/7} E_{55}^{3/7}
t_2^{-2/7} ~{\rm eV}.
\label{Emax_ra_i}
\ee
\be 
E_p^2 J_p(E_p) = 1.6\times 10^{-9} (1+z)^{4/7} \xi_1^{-1} \eps_p
n_0^{1/7} \Gamma_{2.5}^{-6/7} E_{55}^{6/7} t_2^{-4/7} d_{28}^{-2}
~{\rm GeV~cm}^{-2}~{\rm s}^{-1}.
\label{Jp_ra_i}
\ee
%

\subsection{Adiabatic blastwave in wind}

\ba
E_{p,l} &=&  9.1\times 10^{7} (1+z)^{-2} \eps_{B,-1}^{-1/2} 
\eps_{e,-1}^{-2} A_\star^{-1/2} t_2 ~{\rm GeV},
\nonumber \\
E_{p,h} &=&  3.2\times 10^{12} \eps_{B,-1}^{3/2} 
A_\star^{3/2} t_2^{-1} ~{\rm GeV}.
\label{CRbreaks_ad_w}
\ea
\be
\tau_{p\g} (E_{p,l}) = 6.0\, (1+z)^{1/2} \eps_{B,-1}^{1/2}
A_\star^{2} E_{55}^{-1/2} t_2^{-1/2}.
\label{opt_ad_w}
\ee
\be
{\cal E}_{CR} = \eps_p E_k/2.
\label{totCR_ad_w}
\ee
\be
E_{p,s} =  6\times 10^{18} (1+z)^{-5/4} \phi_1^{-1} A_\star^{-1/4} 
\eps_{B,-1}^{1/2} E_{55}^{3/4} t_2^{1/4} ~{\rm eV}.
\label{Emax_ad_w}
\ee
\be 
E_p^2 J_p(E_p) = 1.2\times 10^{-8} (1+z)^{1/2} \xi_1^{-1} \eps_p
A_\star^{1/2} E_{55}^{1/2} t_2^{-1/2} d_{28}^{-2} ~{\rm
  GeV~cm}^{-2}~{\rm s}^{-1}.
\label{Jp_ad_w}
\ee
%

\subsection{Radiative blastwave in wind}

\ba
E_{p,l} &=&  1.6\times 10^{8} (1+z)^{-2} \eps_{B,-1}^{-1/2} 
\eps_{e,-1}^{-2} A_\star^{-1/2} t_2 ~{\rm GeV},
\nonumber \\
E_{p,h} &=&  6.0\times 10^{11} \eps_{B,-1}^{3/2} 
A_\star^{3/2} t_2^{-1} ~{\rm GeV}.
\label{CRbreaks_ra_w}
\ea
\be
\tau_{p\g} (E_{p,l}) = 12.6\, (1+z)^{1/3} \eps_{B,-1}^{1/2}
A_\star^{13/6} \Gamma_{2.5}^{2/3} E_{55}^{-2/3} t_2^{-1/3}.
\label{opt_ra_w}
\ee
\be
{\cal E}_{CR} = 6.4\times 10^{53} (1+z)^{1/3} \eps_p A_\star^{1/3} 
\Gamma_{2.5}^{-4/3} E_{55}^{4/3} t_2^{-1/3} ~{\rm erg}.
\label{totCR_ra_w}
\ee
\be 
E_{p,s} = 8\times 10^{17} (1+z)^{-1} \phi_1^{-1} \eps_{B,-1}^{1/2}
A_\star^{-1/2} \Gamma_{2.5}^{-1} E_{55} ~{\rm eV}.
\label{Emax_ra_w}
\ee
\be 
E_p^2 J_p(E_p) = 3.3\times 10^{-9} (1+z)^{2/3} \xi_1^{-1} \eps_p
A_\star^{1/3} \Gamma_{2.5}^{-2/3} E_{55}^{2/3} t_2^{-2/3} d_{28}^{-2}
~{\rm GeV~cm}^{-2}~{\rm s}^{-1}.
\label{Jp_ra_w}
\ee
%

\section{Pion and muon decay scaling functions}
\label{appendC}

For completeness, we quote here the spectra of secondary particles,
called scaling functions, for the pion and muon decays given in
Ref.~\cite{Lipari:1993hd}.  In terms of the ratio between the muon and
pion masses squared: $r_\pi = m_\mu^2/m_\pi^2$, the pion decay scaling
relations are
\ba
f_{\pi \to \mu} (x) &=& \frac{1}{1-r_\pi} \Theta (x-r_\pi) ,
\nonumber \\
f_{\pi \to \nu_\mu} (x) &=& \frac{1}{1-r_\pi} \Theta (1-r_\pi-x).
\label{pi_mu_dist}
\ea
Note that pion decay muons are polarized and one should take into
account their helicities (negative for $\mu^+$ and positive for
$\mu^-$, on the average) since the neutrino spectra from muon decay
depend on the polarization.  The helicity function, in case of
ultra-relativistic pion decay, is given by
\ba
P_{\pi^\pm \to \mu^\pm}(x) = \pm \frac{2r_\pi}{x(1-r_\pi)} \mp
\frac{1+r_\pi}{1-r_\pi} .
\label{helicity}
\ea
The muon decay scaling relations are
\ba
f_{\mu^+ \to \nu_e} (x,y) 
&=& (2 - 6y^2 +4y^3) + P_{\pi^+\to\mu^+}(x) (-2+12y-18y^2+8y^3),
\nonumber \\
f_{\mu^+ \to {\bar \nu}_\mu} (x,y)
&=& \left(\frac{5}{3} - 3y^2 + \frac{4}{3}y^3 \right) 
+ P_{\pi^+\to\mu^+}(x) \left(\frac{1}{3} - 3y^2 + \frac{8}{3}y^3 
\right).
\label{mu_nu_dist}
\ea

\end{widetext}

\end{document}